\documentclass[twocolumn,showpacs,showkeys,preprintnumbers,amsmath,amssymb,aps]{revtex4}

\usepackage{graphicx}
\usepackage{dcolumn}
\usepackage{bm}
\usepackage{natbib}
\usepackage{color}
\usepackage{nomencl}
\usepackage[ansinew]{inputenc}
\usepackage{rotating} 

\newcommand{\unit}[1]{\ensuremath \,{\rm{#1}}}
\usepackage{upgreek}

\begin{document}

\preprint{Physical Review ST AB preprint}

\title{On the Response of an OST to a Point-like Heat Source}

\author{A. Quadt}
\author{B. Schr\"{o}der}
\author{M. Uhrmacher}\email{Michael.Uhrmacher@gmx.de}
\author{J. Weingarten}
\author{B. Willenberg}
\author{H. Vennekate}

\affiliation{Universit\"{a}t G\"{o}ttingen, II.~Physikalisches
Institut, Friedrich-Hund-Platz~1, 37077~G\"{o}ttingen, Germany}

\date{\today}

\begin{abstract}
A new technique of superconducting cavity diagnostics has been introduced by D. Hartrill at Cornell University, Ithaca, USA. Oscillating Superleak Transducers (OST) detect the heat transferred from a cavity's quench point via {"}Second Sound{"} through the superfluid He bath, needed to cool the superconducting cavity. The observed response of an OST is a complex, but reproducible pattern of oscillations. A small helium evaporation cryostat was built which allows the investigation of the response of an OST in greater detail. The distance between a point-like electrical heater and the OST can be varied. The OST can be mounted either parallel or perpendicular to the plate, housing the heat source. If the artificial quench-point releases an amount of energy compatible to a real quench spot on a cavity's surface, the OST signal starts with a negative pulse, which is usually strong enough to allow automatic detection. Furthermore, the reflection of the Second Sound on the wall is observed. A reflection coefficient $R = 0.39 \pm 0.05$ of the glass wall is measured. This excludes a strong influence of multiple reflections in the complex OST response. Fourier analyses show three main frequencies, found in all OST spectra. They can be interpreted as modes of an oscillating circular membrane. 
\end{abstract}

\pacs{03.75.Kk, 07.20.Mc, 43.38.+n, 67.25.dm, 67.25.df, 67.25.du}
                            
\keywords{Second Sound, cavity quench, Oscillating Superleak Transducer (OST), superconducting cavity, cavity diagnostics, He evaporation cryostat, ILC}
\maketitle

\section{Introduction}
\label{sec:introduction}
In modern particle accelerators the cavities are usually made of superconducting materials like niobium with a high critical temperature of $T_{c} = 9.2\unit{K}$.  At the low operating temperature of a cavity (1.5 - 4~K) the Ohmic losses are clearly reduced and almost all the RF power from the Klystrons is available to accelerate the beam. Quality factors of $Q_{0} = 10^9 - 10^{10}$ are obtained. Niobium can be sputtered on the interior surface of a copper cavity like in the LHC, or the whole cavity can be made from ultra-pure niobium. A typical example of the latter one is the so called TESLA cavity \cite{1,2} which consists of 9 cells, has a total length of about $1\unit{m}$ and is operated at $1.3\unit{GHz}$. These cavities are used at FLASH and soon at the XFEL, both free electron lasers at DESY, and they are considered for the future ILC collider.

The niobium coated copper cells of the LHC produce an accelerating electric field gradient of $5\unit{MV/m}$. The TESLA cavities were designed to run at a gradient of at least $15\unit{MV/m}$. They should reach $23.6\unit{MV/m}$ at the XFEL and up to $31.5\unit{MV/m}$ for the ILC \cite{3,4}. Still higher gradients would reduce the number of cavities needed and lower the costs. But as the accelerating electric field is linked to the magnetic field, the maximum gradient is limited by the critical magnetic field $H_c$. According to current knowledge \cite{5}, niobium with $H_{c}\sim 200\unit{mT}$ at 2~K is probably useable for gradients up to  $E_{acc}\sim 55 - 60\unit{MV/m}$.

There are typical problems to reach these high gradients. Electrons and discharges are produced inside the cavity, which often start at imperfections of the inner walls. Heat is generated at these spots, the superconductivity is lost and still more heat occurs due to resistive heating: the cavity quenches and the stored energy is released. To avoid this, during the course of cavity-development several different treatments are used (grinding, high pressure water rinsing, heat treatment, buffered chemical or electro polishing, etc.) \cite{2} to optimize the inner surfaces of the cavities.

Nevertheless, each cavity has to be tested, before it is inserted into a larger accelerator facility. The present 500\unit{GeV} ILC design requires about 18.000 TESLA niobium cavities (including spares) yielding an average operational field gradient of $E_{acc}= 31.5\unit{MV/m}$ \cite{6}. The classical way of testing a cavity for quench spots is the use of arrays of thermocouples, which are moved around the cavity in operation. Quench spots can be localized to within $2-3\unit{cm}$, but \textit{``the process of locating defects in 9 cell cavities remains a lengthy and cumbersome process''} \cite{7}. It can hardly be applied on the thousands of ILC cavities.

In 2008, a new technique was introduced at Cornell university to trace the position of a quench spot \cite{8,9}. The principle of the method is the following: When a superconducting cavity is operated below $2.17\unit{K}$, the liquid He in the cooling container becomes superfluid (He-II). In case the cavity cannot stand the accelerating gradient, a quench occurs. The temperature at the quench point rises and generates a so called Second Sound wave which travels through the liquid He-II. The Oscillating Superleak Transducer (OST) can be used to track such a wave in this environment. The time interval between the quench and the arrival of the Second Sound at an OST can be measured, typically 0.5~ms per cm distance. The application of multiple OSTs at selected positions allows  a localization of the quench spot via triangulation.

The signal of an OST is a complex superposition of the Second Sound wave and the eigenmodes of the OST membrane. Further complication of the signal shape can be generated by Second Sound waves  reflected at the cavity walls, the walls of the cryostat, etc. To disentangle these effects, the response of an OST is studied in the simplified environment of a small He evaporation glass cryostat.

\section{Second Sound in Superfluid Helium}
\label{sec:SeSo}

\subsection{Properties of Liquid He}
\begin{figure}[ht!]
	\vspace{1em}
	\centering
	\includegraphics[]{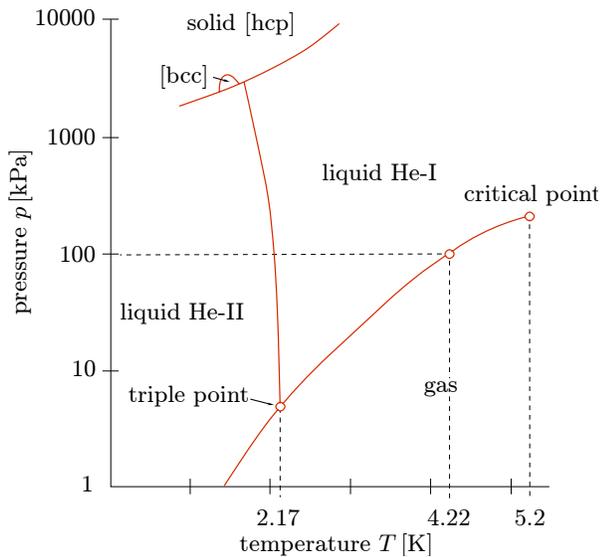}
	\caption{Helium phase diagram \cite{9}. Cooling down helium under normal pressure results at $4.2\unit{K}$ in the liquid phase. Reduction of the pressure to $49\unit{hPa}$ cools the liquid to the ``Triple-point'' where superfluidity occurs.}
	\label{fig:HePhasediagram}
\end{figure}

Cavities in a modern accelerator are placed inside a cooling bath of liquid He. The phase diagram (Fig.~\ref{fig:HePhasediagram}) shows, that at normal pressure ($100\unit{kPa}$) liquid He has a temperature of $4.22\unit{K}$. If the pressure is reduced the He cools down to the Triple point $T_{\lambda} = 2.17\unit{K}$ at the pressure of $p_{\lambda} = 4.9\unit{kPa}$ \cite{10}. Below the so called  $\lambda$-point liquid He undergoes a transition into a superfluid phase (He-II), where it has no internal friction and nearly perfect heat conductivity.

A phenomenological description of this transition was first given by Tisza and later by Landau \cite{11}. They suggested an ideal mixture of two fluids below $T = 2.17\unit{K}$: a normal- and a superfluid fraction. The normalfluid fraction has the density ${\rho_n}$ and the viscosity ${\eta_n}$. It travels with the velocity $v_{n}$ and carries the total entropy $S$ of the system. The superfluid fraction has the density $v_{s}$, a viscosity $\eta_s = 0$, a vanishing entropy and travels with the velocity $v_{s}$. The total density is the sum of the two components:  $\rho = {\rho_s} + {\rho_n}$.

\begin{figure}[ht!]
	\vspace{1em}
	\centering
	\includegraphics[]{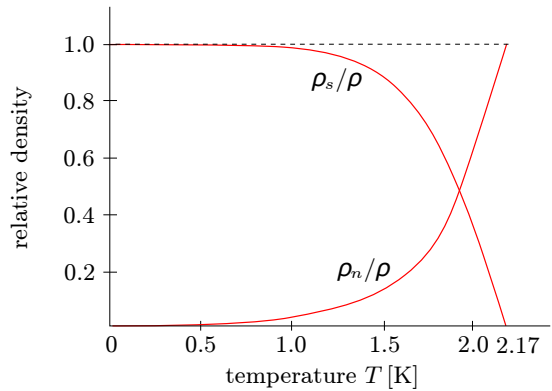}
	\caption{Relative density versus temperature below the $\lambda$-point in helium.}
	\label{fig:HeRelativeDensity}
\end{figure}

As shown in Fig.~\ref{fig:HeRelativeDensity} the density ratio is strongly temperature dependent between $1.0$  and $2.17\unit{K}$. One can define $j$, the mass flow of He below the $\lambda$-point to be: $j = \rho_s  v_s + \rho_n  v_n$. As shown in \cite{12,13}, now two types of heat transport in liquid He below the $\lambda$-point can be derived: The First Sound, a pressure wave with velocity ${v_{FiSo}}$, and the Second Sound, an entropy wave with velocity ${v_{SeSo}}$.

\subsection{Second Sound}
While the First Sound is a longitudinal fluctuation in density, driven by pressure, the Second Sound is a temperature driven wave, opposite in phase movement of the normal and superfluid components - it is an entropy wave that transports heat. The velocity $v_{SeSo}$  of this heat transfer has a strong temperature dependence, as can be deduced from the temperature dependence of the density ratios given in fig.~\ref{fig:HeRelativeDensity}. Various measurements were performed to determine $v_{SeSo}$.

\begin{figure}[ht!]
	\vspace{1em}
	\centering
	\includegraphics[]{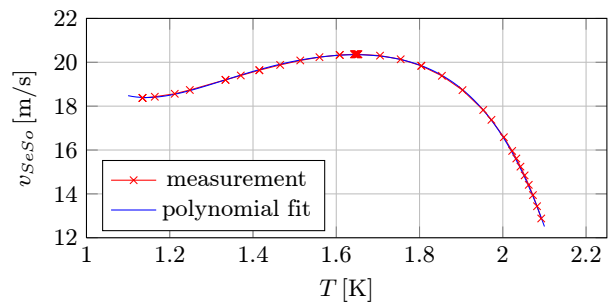}
		\caption{Second Sound velocity  $v_{SeSo}$ versus temperature \cite{14}.}
		\label{fig:SeSoVelo}
\end{figure}

Fig.~\ref{fig:SeSoVelo} shows data from the measurement of $v_{SeSo}$ performed by Wang et al. \cite{14} in resonance experiments. The blue curve is a polynomial fit to the data of \cite{14}. During the experiments presented here, the temperature is kept between $1.5$ to $1.8\unit{K}$, where the velocity of the Second Sound is nearly constant at a value of  $v_{SeSo}\sim 20\unit{m/s} = 2\unit{cm/ms}$. Nevertheless, the actual velocity is derived from the measured temperature of the He using the polynomial fit to Wang`s \cite{14} data.

\section{Second Sound Ray-tracing around a Tesla Cavity -- a Simulation}
\label{sec:Rays}

\subsection{Simplifying Assumptions}
The Second Sound localisation technique is studied using simulations of the TESLA  nine-cell cavity. The dimensions of the cavity are taken from \cite{1}. The length of the whole structure is $1276\unit{mm}$, while the single cells are  $115.4\unit{mm}$ long and $206\unit{mm}$ wide. The diameter of the iris diaphragms between the cells is $70\unit{mm}$. In the simulation program the shape is simplified to a central cylinder and nine half-tori, which cover the critical regions of the cavity well. All additional parts like HOM couplers are neglected. The critical region is the equator of each cell where two half-cells are welded together. The resulting welding seam is the main source for material disorder and it coincides with the region of highest magnetic field. Therefore, for the simulation, the quench spots are randomly placed around the equators of the cells.

\begin{figure}[h!]
	\vspace{1em}
	\centering
	\includegraphics[]{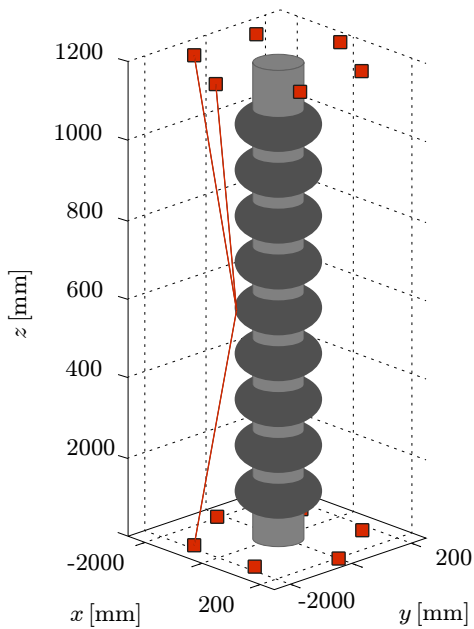}
	\caption{TESLA cavity in the simulation with  $2 \times 6$ OSTs (red squares). Shown is a three signal event.}
	\label{fig:SimulationTeslaCavity}
\end{figure}

The geometry of the liquid He tank, which is planned to be used at DESY for testing the TESLA cavities, limites the placement of the OSTs. To allow for a fast cavity exchange during testing, one plane with OSTs can be mounted permanently on the bottom of the tank, the others can be mounted on the cover of the tank. The distance between OST and equator is assumed to be $103\unit{mm}$, i.e. on a circle with twice the maximum radius of the cavity.

A constant velocity of the Second Sound is assumed ($v = 20\unit{m/s}$) in the simulations. For the reconstruction only lines of direct sight are used. As a consequence, every quench above a cells equator should not be seen by a bottom detector and vice versa. According to the wave-nature of Second Sound, events within a distance of less than 20~mm distance to the cells equator are propagated around the equator to be registered by an OST on the other side. It was shown that already two planes with six detectors are sufficient in the selected geometry to have always two OSTs with direct lines of sight to the quench spot. No reflections and no attenuations are implemented. Details of the simulations can be found in \cite{15}.

\subsection{Results and Consequences}
As a first result, the simulations show that in the chosen case of six detectors on each of the two levels (on the top and bottom of the cryostat) quenches are in general detected by two or three OSTs with direct lines of sight. 
Two-signal events, which are predominantly observed for quenches in the central cells can be divided in two sub-groups. When the two firing OSTs sit on the same plane  the reconstruction is straight-forward. The critical case is a signal in two OSTs in different planes (see Fig.~\ref{fig:SimulationTeslaCavity} and take only one red line in the upper part). The reconstruction will give two possible quench locations. In most cases, one of the proposed locations is within a few millimeters from the original quench spot. The mean deviation between a real spot and the reconstructed one is about $5.7\unit{mm}$.

In the case of three or more responding OSTs (see Fig.~\ref{fig:SimulationTeslaCavity}) the spot is unambiguously localized with a mean deviation of only $4.5\unit{mm}$. A third plane of OSTs in the middle of the cryostat, i.e. of the cavity, would increase the localisation precision. The simulations show, that it is possible to cover nearly $99\unit{\%}$ of the critical regions of a TESLA cavity with twelve detectors on two levels, six at the top and six at the bottom of the cryostat \cite{15}.
Nevertheless, the geometry of a real cavity is more complex and additional big parts might be inside the He tank. The question arises whether strong reflections of the Second Sound on the surfaces occur. These would increase the complexity of the quench-localisation algorithms.

\section{Experiments inside a Helium Evaporation Cryostat}
\label{sec:experiments}

\subsection{The apparatus}

\begin{figure}[!ht]
	\vspace{1em}
	\centering
	\includegraphics[]{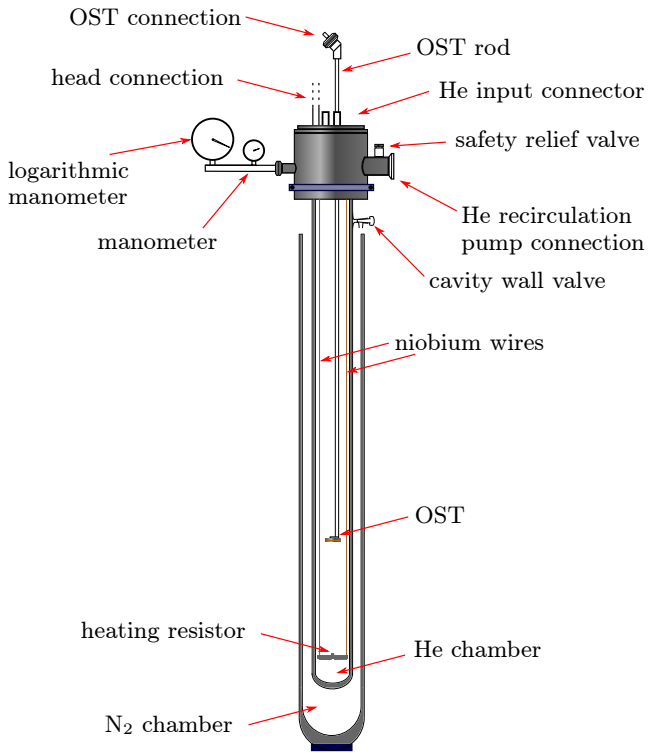}
	\caption{Sketch of the He evaporation cryostat with mounted OST and heat source.}
	\label{fig:figure5}
\end{figure}

A He evaporation cryostat was built as a test volume for Second Sound waves, produced by a tiny heat source and measured by an OST which is mounted on a movable support rod \cite{16}. The cryostat consists of two nested Dewars. The outer one contains liquid $N_2$ as a cooling shield for the inner Dewar, filled with liquid He. The hollow glass walls are evacuated and mirrored to provide a high degree of heat insulation. The inner He Dewar is closed with a stainless steel chamber, which houses all necessary connections (electrical for OST and heater, mechanical for vacuum pumping, pressure measurements, a feed-through for the OST mount and one for filling the liquid He into the inner Dewar). The heat source (see section \ref{subsec:reflections}) is positioned on the central axis close to the bottom of the Dewar. The OST rod is pivoted and can be positioned at any height in the He Dewar. The rod axis is shifted by $12.5\unit{mm}$ with respect to the central axis of the 70~mm wide Dewar to allow for noncentral positioning of the OST by turning. Fig.~\ref{fig:figure5} shows the setup. A rotary vane pump (Alcatel 2033 with $30\unit{m^3/h}$ suction power) is used to lower the pressure in the gas volume above the surface of the liquid He. Evaporation cools the liquid to $1.5$ or $1.8\unit{K}$ (approximately $5 - 17\unit{hPa}$) into the superfluid He phase. By pumping out the He vapour the total amount of He in the Dewar is reduced over time. A maximum distance between the heat source and the OST of about $44\unit{cm}$ can be realized at the beginning of the measurements. After 8 hours the height of the liquid He is still $20\unit{cm}$.
The temperature is controlled by the pressure of the gaseous He according to the International Temperature Scale (ITS-90) \cite{17}. The precision of the temperature determination is obtained from the pressure measurement. In the experiment we used a logarithmic membrane vacuum meter (DIAVAC DV1000, Oerlikon Leybold Vacuum), connected by a steel bellow to the cryostat's head. The absolute accuracy in the pressure range of $1$ to $10\unit{hPa}$ is claimed to be $\pm1\unit{hPa}$ \cite{18}. This corresponds to an error of the temperature of $\Delta T = \pm 0.05\unit{K}$.    

\subsection{OST-Design and Working Principle}
\label{subsec:ostDesign}

\begin{figure}[!ht]
	\vspace{1em}
	\centering
	\includegraphics[]{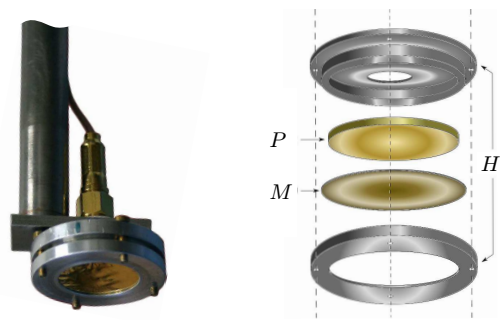}
	\caption{OST mounted horizontally on the support-tube (left) and exploded-drawing (right) with $H$ = housing, $M$ = porous membrane and $P$ = immobile brass plate.}
	\label{fig:figure6}
\end{figure}

Oscillating Superleak Transducers (OST) can be used for excitation and detection of Second Sound in superfluid helium. The concept is similar to a condenser microphone. The functional part is a membrane ($M$ in Fig.~\ref{fig:figure6}, active diameter: $19.6\unit{mm}$) made from a Cellulose Nitrate Membrane Filter (VWR Scientific, USA) of $0.1\unit{mm}$ thickness, which contains pores with a diameter of $0.2\unit{\upmu m}$ \cite{16}.

These pores allow the passage of superfluid He, whereas the normalfluid He cannot pass them. The second electrode of the capacitor is realised by an immobile brass plate (P in the figure). The total capacity of the OST is estimated to be $C_{OST} = 16\unit{pF}$. One side of this filter paper M is coated by a thin Au-layer ($50\unit{nm}$). This side is directed towards the arriving Second Sound wave. The OST follows closely the first concept of Cornell \cite{9}, was adapted at DESY \cite{19} and supplied to us.

\begin{figure}[!ht]
	\vspace{1em}
	\centering
	\includegraphics[]{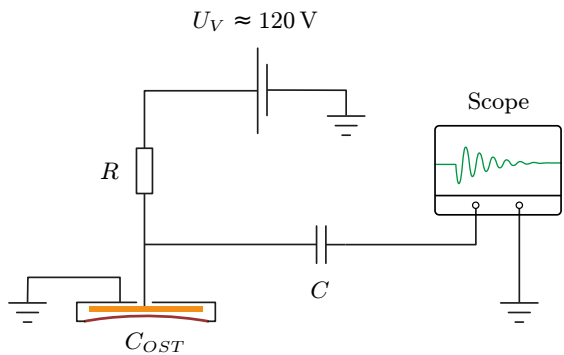}
	\caption{Readout circuit for the OST.}
	\label{fig:figure7}
\end{figure}

Fig.~\ref{fig:figure7} shows the readout circuit for the OST. A battery charges the OST with a voltage of $120\unit{V}$. In the first experiments a resistance of $R = (100 \pm 5)$~k$\Omega$ and a decoupling capacity of $C = (100 \pm 5)\unit{\upmu F}$ were used, later these elements were exchanged to optimise the signal amplitude. The best signals are obtained with $R = (10 \pm 0.5)\unit{M\Omega}$ and $C = (22 \pm 2)\unit{pF}$.

The response of the OST to the arriving Second Sound wave depends on the density ratio $\rho_n$ / $\rho_s$. The heat pulse from a quench spot (or from our point-like heat source) increases $\rho_n$, the density of normalfluid He. When the entropy wave reaches the OST, He II diffuses from the inner volume, the membrane moves inward increasing the capacity $C_{OST}$. As the stored charge $Q = C_{OST} \cdot U$ is stable, the voltage over the capacitor is lowered creating a negative pulse on the oscilloscope. In the present experiments, a Textronix TDS 4014 B is used in ``high resolution mode'' to detect the OST signal with lower noise.

\subsection{Heat Source}
The point-like heat source is realised by a single $56\unit{\Omega}$ SMD-resistor with the dimensions 
$1.5 \times 2.2\unit{mm^2}$. The resistor is placed on a PVC plate. This plate is suspended from three niobium wires (diameter: $0.5\unit{mm}$). Weights are mounted below the plate to damp mechanical oscillations, should they arise. Two of the three wires carry the current-pulse to the resistor. Niobium is selected for the wires, as it is superconducting below $9.2\unit{K}$. Therefore no Ohmic heating occurs, which could be an additional source of Second Sound.

To work under realistic conditions, the power of the heat source should correspond to the heat power  $P_{diss}$ which is dissipated per area A, when a quench occurs. According to \cite{5}, one can calculate:       
\begin{align}
P_{diss} = \pi/2\cdot r^2 \cdot R_A \cdot H^2 .
\end{align}

We assume a circular defect of the radius $r = 250\unit{\upmu m}$. In case of a 9 cell TESLA cavity, operated at a frequency of $1.3\unit{GHz}$, the magnetic field in the equatorial regions is in the order of $H \approx 100\unit{Oe}$ and the surface resistance $R_A$ rises during the quench to a few m$\Omega$ \, assuming a quenchtime of abaout $5\unit{ms}$ \cite{8}. Under these assumptions one estimates a dissipated energy of $0.15\unit{mJ}$ and a dissipateted power $P_{diss}$ of $0.26\unit{mW}$ \cite{5}. In the experiments different values for voltage and pulse length are used to study the influence of the heat-pulse shape on the OST response (see Fig.~\ref{fig:figure9}). At the temperature of $2\unit{K}$, the resistance of the heat source is measured to be $R = (82.5 \pm 1.0)\unit{\Omega}$. A rectangular pulse of $0.2\unit{ms}$ duration and an amplitude of $4\unit{V}$, or with a length of $0.8\unit{ms}$ and an amplitude of $2\unit{V}$ is applied. Both pulses deliver $(0.039 \pm 0.001)\unit{mJ}$. The best response of the OST is found when higher energy is generated in the pulse: with an amplitude of $4\unit{V}$ and a pulse duration of $0.8\unit{ms}$, an energy of $(0.156 \pm 0.002)\unit{mJ}$ is dissipated.

\section{Measurements}
\label{sec:measurements}

\subsection{Noise Suppression}
In the first experiments, a high electronic background of $50\unit{Hz}$ and higher order modes made it difficult to identify the OST response to a heat pulse of maximal $2\unit{mV}$ amplitude. A consequent grounding, the application of a battery instead of a variable power supply and the change of the resistor $R$ and the decoupling capacity $C$ (see Fig.~\ref{fig:figure9}) increased the signal/noise ratio to some extend.  As the next step the length of the heating pulse was extended to $0.8\unit{ms}$. The increased heat-power caused larger OST amplitudes.

The main source of noise is the rotary vane pump. The OST - sensitive to the First Sound, too - detects $50\unit{Hz}$ oscillations caused by the running pump. Switching off this pump, one obtains a nearly flat background within $10\unit{s}$. Nevertheless, switching off the pump has the consequence, that the temperature of the liquid He increases, changing  the velocity $v_{SeSo}$ too. Therefore, the measurements are performed in the temperature interval from $1.5$ to $1.8\unit{K}$, which corresponds to a pressure range from $5$ to $17\unit{hPa}$. According to \cite{13} and shown in Fig.~\ref{fig:SeSoVelo} here the velocity $v_{SeSo}$ is nearly constant with $1.987$ to $2.004\unit{cm/ms}$, a  variation of only $1\unit{\%}$ of the minimal value. These conditions determine the scheme of the measurements: After a  time of about $15\unit{min}$, the He vapour reaches $17\unit{hPa}$. Data taking is stopped, the pump is switched on to reduce the pressure again to $5\unit{hPa}$ within the next $15\unit{min}$. Then new measurements can be started. 

\subsection{Travelling Distance Variation}

\begin{figure*}
	\centering
	\includegraphics[]{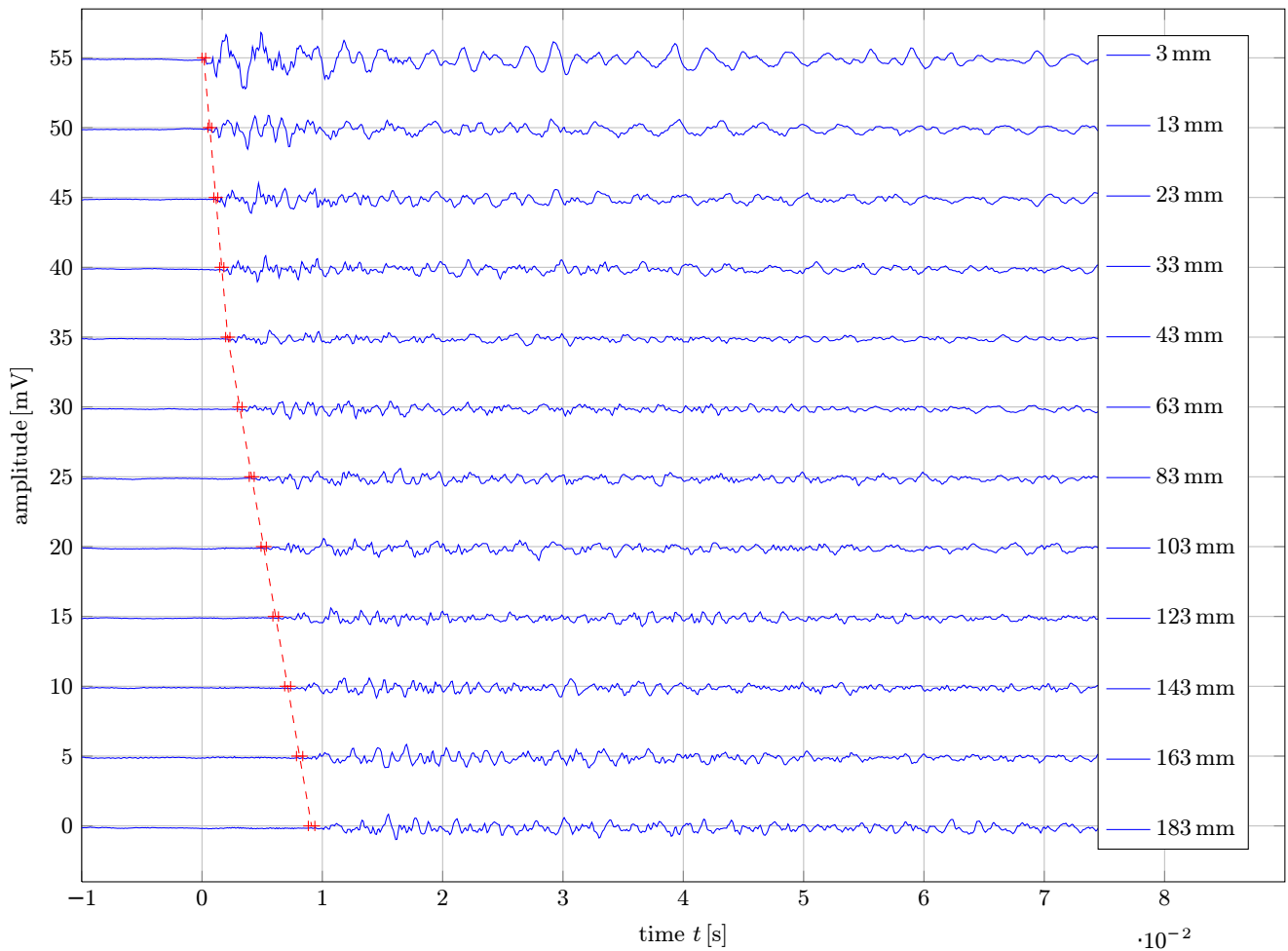}
	\caption{12 OST signals recorded in the high resolution mode with increasing distance (given
               on the right side of each signal). The He level was $19.5\unit{cm}$ above the resistor. The red
               crosses mark the earliest and latest expected point for the Second Sound to reach
              the OST; the kink in the red dotted line is due to the change in the step size. All 12 signals are shown with the same scale for the amplitude: $-5\unit{mV}$ to $+5\unit{mV}$. }
	\label{fig:figure8}
\end{figure*}

The primary goal is the exact determination of the arrival time of a Second Sound wave. The distance between the heat source and the OST is adjusted in well defined steps. The zero distance is experimentally determined by mechanically contacting the OST and the heat source. Nevertheless, uncertainties in the distance remain due to the thermal expansion of the used materials. The total error in the distance $z_{OST}$ (OST/heat source) is estimated to be $\pm 3\unit{mm}$ \cite{16}. Fig.~\ref{fig:figure8} gives an overview of 12 of the signals recorded with increasing distance $z_{OST}$, indicated on the right axis of the figure.

The heat pulse is released for all 12 spectra at $t = 0\unit{ms}$. To each line two red crosses are added marking the earliest and latest point in time when the Second Sound signal is supposed to reach the OST, calculated from the actual pressure and distance. The time interval between each pair is the sum of the deviation of the rising and falling edge of the trigger signal and the uncertainty in the Second Sound velocity in the actual temperature interval. The first crosses are connected by a dotted red line. The kink in this line at $z_{OST} = 43\unit{mm}$ is due to the change in the step size in  $z_{OST}$  from $10$ to $20\unit{mm}$. 

Fig.~\ref{fig:figure8} shows a complex but reproducible pattern of oscillations, which is different for each selected distance. The signals start with a negative deflection, but with a reduced amplitude. OST signals from the shortest distance (3~mm) show the highest amplitudes. The amplitudes decrease up to a distance of $40\unit{mm}$  and then stay at a constant value.

\begin{figure}[!ht]
	\centering
	\includegraphics[]{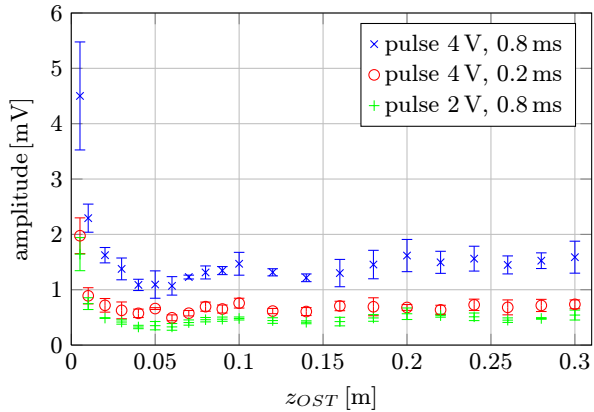}
	\caption{Signal amplitude at the OST depending on the travelling path $z_{OST}$ of the Second Sound wave. Three different 
	      	 shapes of the heating pulse are compared.}
	\label{fig:figure9}
\end{figure}

This behaviour is clearly seen for the 0.8~ms long 4~V heat pulses in Fig.~\ref{fig:figure9} and reflects the fact, that the Second Sound propagates as a spherical wave close to the heat source. From a distance of about $40\unit{mm}$ on, it can approximately be regarded as a planar wave with constant amplitude. The complicated pattern of the OST response raised the important question on the reproducibility of these patterns. Therefore we recorded at least 10 OST spectra under identical conditions at each distance $z_{OST}$. All signal patterns from the same distance are equal, but different for another value of $z_{OST}$.

\begin{figure}[!ht]
	\centering
	\includegraphics[]{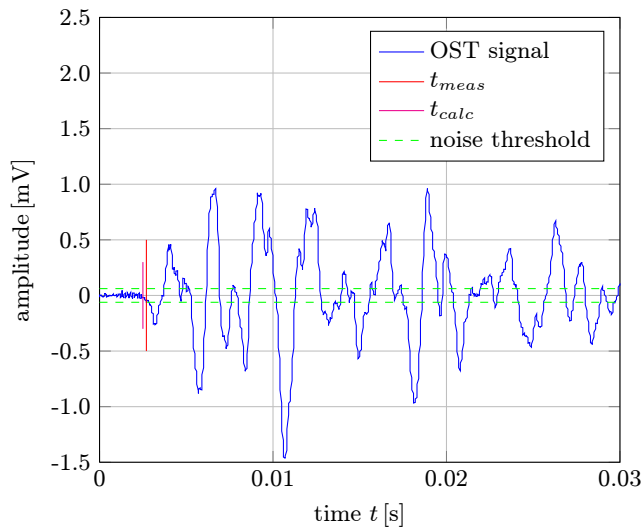}
	\caption{Starting region of an OST signal. In green the noise band is given, calculated from the 
               measured OST-signal prior firing the resistor. The first red line ($t_{calc}$) indicates the position,
               where the signal is expected theoretically, the second one ($t_{meas}$) gives the position
               where at first the OST signal crosses the noise band.}
	\label{fig:figure10}
\end{figure}

As already discussed in section~\ref{subsec:ostDesign}, the OST should respond with a negative pulse to the arrival of a wave front of normalfluid He, i.e. the Second Sound wave from the heat source. The exact determination of this arrival time is  crucial for the localisation of quench points. As the signal does not start with the full amplitude - a slow onset of membrane oscillations occurs (see section~\ref{subsec:eigenmodes}) - it is difficult to separate the small negative pulse from the noise background. The maximum amplitude of this noise band can be determined from the data, taken in the time interval prior to the arrival of the Second Sound wave at the OST. The threshold is placed $10\unit{\%}$ higher than the maximal noise amplitude (green lines in fig.~\ref{fig:figure10}). The determined arrival time $t_{meas}$ is indicated together with the arrival time $t_{calc}$, calculated from the known OST distance and the actual velocity $v_{SeSo}$. The agreement is satisfying.

\begin{figure}[!ht]
	\centering
	\includegraphics[]{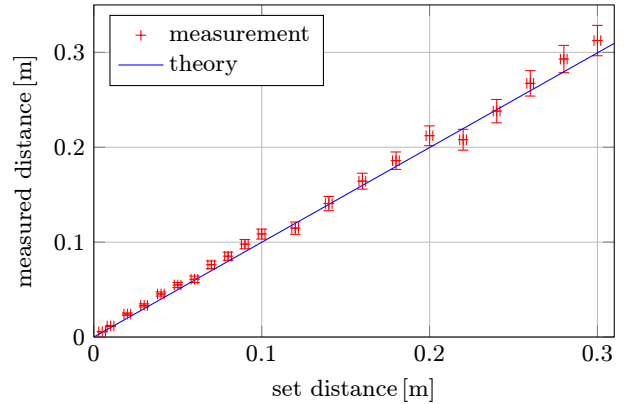}
	\caption{ Distances obtained from data analysis of a full series of OST spectra versus the distance obtained from the mechanical set up.}
	\label{fig:figure11}
\end{figure}

In a full series of OST spectra (obtained with $4/unit{V}, 0.8/unit{ms}$ heat pulses) taken at different distances $z_{OST}$, the arrival time $t_{meas}$ is determined with a computer algorithm. In Fig.~\ref{fig:figure11} the resulting distances (red data points with errors) are plotted versus the set-distance $z_{OST}$. The linear correlation corresponds to the calculated velocity of the Second Sound. The deviation between the measured distance and the set-distance $z_{OST}$ is $6.0\unit{mm}$ \cite{20}.

\subsection{Reflections}
\label{subsec:reflections}
\begin{figure}[!ht]
	\centering
	\includegraphics[scale=0.85]{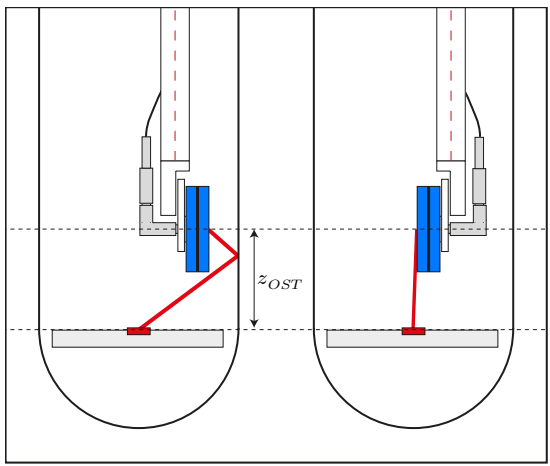}
	\caption{Drawing of the vertical mounting of the OST. In red are shown the shortest distances for the Second Sound to reach the OST with and without reflection. The heat source is plotted in scale.}
	\label{fig:figure12}
\end{figure}

It is important to test whether reflections of the Second Sound can be observed or not. As a positive consequence, reflections would allow to have a look "without direct sight", on the other hand they could be the origin of the complex shape of the OST's signal. For this investigation, the OST is mounted vertically on the supporting rod, which is off-axis (see IV.A). As shown in Fig.~\ref{fig:figure12}~(left), the Second Sound wave can reach the OST only via a reflection on the Dewar glass wall, when the OST is directed towards the wall. A turn of the OST by $180^\circ$ (Fig.~\ref{fig:figure12}, right) allows for a direct impact of the Second Sound wave on the OST. 

\begin{figure}[!h]
	\centering
	\includegraphics[]{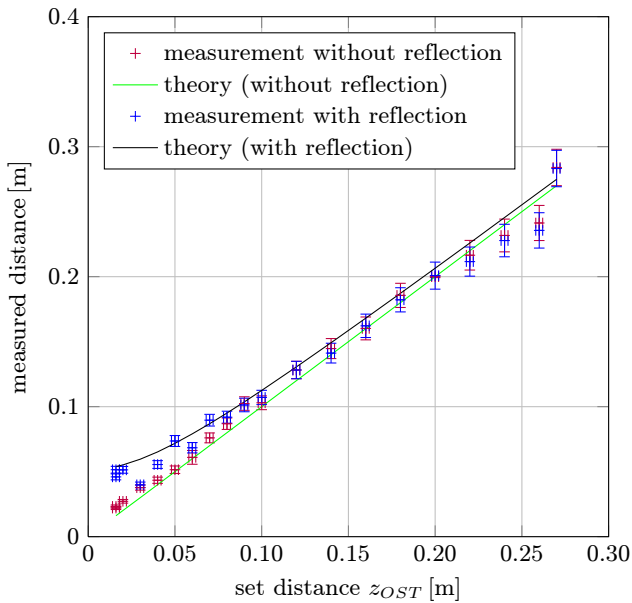}
	\caption{Without reflection (red crosses) the pathlength $P_{SoSe}$ follows the expected linear behavior (see also Fig.~\ref{fig:figure10}). Reflected Second Sound (black crosses) has a longer pathlength $P_{SoSe}$, according to
	         the distance to the wall. The difference in the pathlength's $P_{SoSe}$ shrinks with increasing $z_{OST}$, as the 
	          direct and reflected pathlength gets more similar.}
	\label{fig:figure13}
\end{figure}

The geometry is known and for the two positions a different signal arrival time at the OST is expected. The result of a series of measurements is shown in Fig.~\ref{fig:figure13}. For 19 different values of $z_{OST}$, the travelling distance of the Second Sound wave is measured for the two OST orientations. No other parameters (temperature, pressure, or height of the He-level) are changed. The upper bent line in Fig.~\ref{fig:figure13} shows the calculated path length including a reflection. The measured data from the OST facing the wall are in fact larger and agree with the calculated path length including a reflection. The typical error for the distance is smaller than $10$ or $5\unit{\%}$, as  before \cite{20}.                     

The vertical mounting of the OST gives the possibility to estimate the reflectivity of the wall. The data from the previous experiment are used to compare the signal amplitude from a measurement with and one without reflection. To be independent from any amplitude loss during propagation of the wave, only data with identical travelling distance for both OST orientations (direct impact or after reflection) are compared. This urges, that always $z_{OST}$  has to be enlarged for the direct impact position. 10~pairs of data are evaluated averaging the amplitude of the first four peaks of the signal for both cases and allow to estimate a reflection coefficient of  $R = 0.39 \pm 0.05$ \cite{20}.    

\subsection{Oscillations of the OST membrane}
\label{subsec:eigenmodes}
\begin{figure}[!ht]
	\vspace{1em}
	\centering
	\includegraphics[]{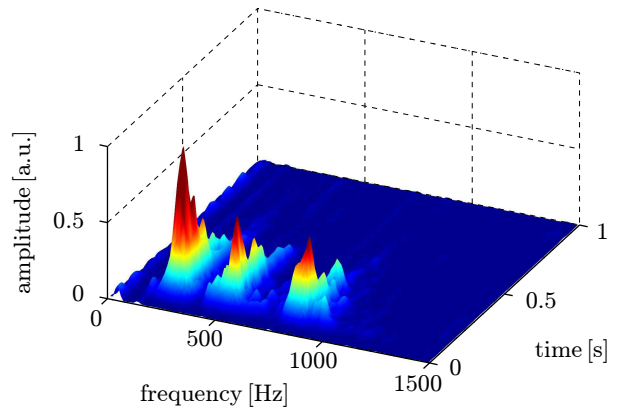}
	\caption{The 3d-diagram shows how the Fourier amplitudes of the three typical frequency distributions decay in time.}
	\label{fig:figure14}
\end{figure}

The complex signal shape of the OST (Fig.~\ref{fig:figure8}) can not be explained by reflections inside the Dewar (see section V.C). Time-resolved Fourier analysis, using a Gaussian window, is used to study the waveform of the OST signal. Fig.~\ref{fig:figure14} shows that three main frequencies contribute to the signal: $f_1 = (299 \pm 5)\unit{Hz}$, $f_2 = (553 \pm 6)\unit{Hz}$ and $f_3 = (870 \pm 13)\unit{Hz}$ \cite{20}. It can be excluded, that these frequencies are caused by the used electronics. Hence, the conclusion is, that these are typical oscillations of the OST, most probably of the OST membrane itself.

\begin{figure}[!ht]
	\vspace{1em}
	\centering
	\includegraphics[]{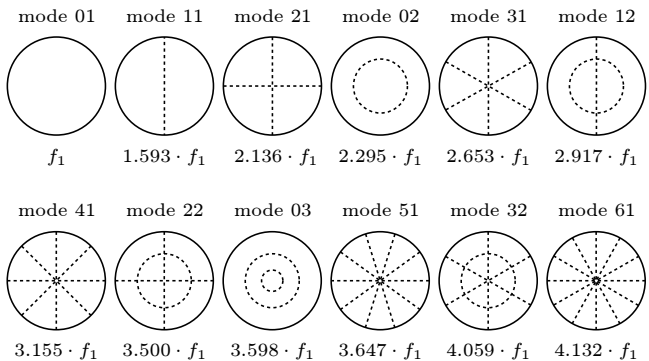}
	\caption{Graphic representation of the eigenmodes with the smallest frequencies of an oscillating  circular membrane, fig. adopted according to \cite{7}.}
	\label{fig:figure15}
\end{figure}

A spherical, flat membrane can oscillate in many modes which are characterized by certain frequency ratios (Fig.~\ref{fig:figure15}) \cite{21}. The ratio of the three main frequencies from the Fourier analysis is $1$:$1.85$:$1.57$. This can be compared to the theoretical resonant frequencies of a circular membrane. Taking the three modes $11, 12$ and $71$ we obtain the ratio  $1$:$1.85$:$1.57$. The maximum deviation to the measured ratio is $0.6\unit{\%}$. If we allow for deviations up to 4 \%, other triplets of modes can reproduce the measured ratios: $(02, 61, 53)$ or $(31, 04, 54)$ or $(32, 21, 53)$. It cannot be excluded that only a part of the OST membrane is oscillating. But it is sure, that at least a part of the membrane is oscillating. This explains the shape of the OST response, the long signal and its exponential decay. To obtain further improvement of the quench localisation, the OST membrane should come into the focus, its shape, thickness and tension.

\subsection{Signal Decay time}
\begin{figure}[!ht]
	\vspace{1em}
	\centering
	\includegraphics[]{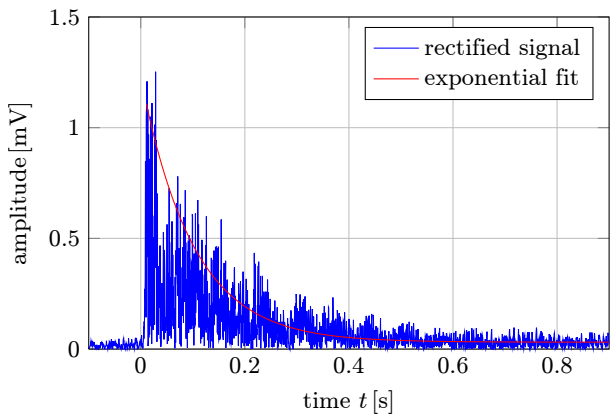}
	\caption{Exponential decay of the OST signal amplitude (absolute value), measured over a full second.}
	\label{fig:figure16}
\end{figure}

The overview (Fig.~\ref{fig:figure8}) already gives the impression, that the OST response in the present setup decays very slowly. The whole range is shown in Fig.~\ref{fig:figure16}, an exponential fit to the absolute amplitudes results in a decay time of  $t \approx (0.14 \pm 0.02)\unit{s}$. One hypothesis suggests multiple reflections in  the small volume of the glass cryostat. But at least 10 reflections are needed to obtain the observed signal length by reflections. The measured  reflection coefficient (see section V.C) excludes this hypothesis, as it would decrease the amplitude down to 1\% of the starting value, whereas the observed value is $1/e \approx 0.37$. The previous section showed, that the OST is a vibratory object which is excited by the incoming Second Sound wavelet. The decay time of the oscillations depends on details of the membranes tension or fixation. OSTs at CERN or DESY have a much shorter signal length \cite{22}, although they are fabricated according to the same design.

\section{Conclusions}
\label{sec:conclusion}
In the present study we show, that the Second Sound wave can still be detected after travelling over a distance of $30\unit{cm}$. The OST signal can be analysed to find the exact arrival time of a Second Sound wave, which is the primary means to localise a quench spot. The complex response signal of an OST in the glass He-evaporation cryostat starts with a "negative" pulse as expected from the working principle. The following oscillations show typical frequencies in certain ratios and an exponential decay of the amplitude: Different modes of an oscillating membrane are observed. 

Additionally, clearly reflections of the Second Sound on the walls of the container are found. The question of their influence on the OST's response is important, especially having in mind the complicated shape of a nine-cell TESLA cavity. The experiments demonstrate a high absorption coefficient at the glass walls, one might expect still higher values on a superconducting niobium metal surface. Therefore, Second Sound waves have only a chance to be detected by an OST after one or two reflections. Furthermore, it justifies the use of direct lines of sight as the most basic way to locate the quench spots. To increase the sensitivity, the amplitudes of the signal have to be enlarged. This could be done by an amplifier or more detailed knowledge on the oscillating membrane. 

\begin{acknowledgments}
We thank Prof. K. Winzer (I. Physikalisches Institut, Georg-August Universit{ä}t G{ö}ttingen) for supplying us with the Dewars for the He-evaporator cryostat and much advice on how to set up the complete cryostat. For valuable discussions about the working principle of an OST we thank E. Elsen, M. Wenskat, F. Schlander and S. Aderhold at DESY. 
The work is supported by the BMBF under contract No. 05H09PX5/HR5/MG5/RD5
\end{acknowledgments}

\bibliographystyle{apsrev}

\end{document}